# Coexistence of different base periodicities in prokaryotic genomes as related to DNA curvature, supercoiling, and transcription

G.I. Kravatskaya*, Y.V. Kravatsky, V.R. Chechetkin, V.G. Tumanyan

*Engelhardt Institute of Molecular Biology of Russian Academy of Sciences, Vavilov str., 32, Moscow, Russia 119991*

___

**Abstract**

We analyzed the periodic patterns in *E. coli* promoters and compared the distributions of the corresponding patterns in promoters and in the complete genome to elucidate their function. Except the three-base periodicity, coincident with that in the coding regions and growing stronger in the region downstream from the transcriptions start (TS), all other salient periodicities are peaked upstream of TS. We found that helical periodicities with the lengths about B-helix pitch ~10.2–10.5 bp and A-helix pitch ~10.8–11.1 bp coexist in the genomic sequences. We mapped the distributions of stretches with A-, B-, and Z- like DNA periodicities onto *E.coli* genome. All three periodicities tend to concentrate within non-coding regions when their intensity becomes stronger and prevail in the promoter sequences. The comparison with available experimental data indicates that promoters with the most pronounced periodicities may be related to the supercoiling-sensitive genes.

___

*Corresponding author: E-mail address:* gk@eimb.ru (G.I. Kravatskaya).

Transcription initiation and gene expression intensity depend on the molecular interactions of transcription factors and RNA-polymerase with regulatory sequences in genomic DNA. These interactions imply not only exact molecular recognition of certain DNA motifs, but also the presence of generic structural features in the corresponding DNA regions. In particular, the static DNA curvature upstream of the transcription start has been shown to modulate transcription [1–4]. Since DNA curvature is related to A/T tracts phased with the pitch of the double helix, the study of A/T patterns with periods 10–11 bp can provide essential information on the regulatory mechanisms [5–9]. In accordance with their regulatory functions, A/T patterns with the helix period tend to concentrate in the intergenic regions [10]. Other A/T periodic patterns in promoter regions, with periods about the spacer length (~17 bp), can also contribute to transcription regulation [11]. The periodicities in purine–pyrimidine distribution along promoter sequences reveal approximate correlations with RNA polymerase–promoter contacts [12]. These examples do not exhaust the important role of DNA periodic patterns in the regulatory mechanisms.

Three-dimensional folding and supercoiling of double-stranded DNA (dsDNA) also affect the transcription [13–16]. Supercoiling is negative in Bacteria. In hyperthermophilic Archaea reverse gyrase induces the positive supercoiling of a chromosome. The negative supercoiling shifts the helical periodicity from ~10.5 bp (free B-form DNA) to ~11 bp, whereas the positive supercoiling shifts it toward ~10 bp [17–19]. Therefore, the helical periodicity in genomic sequences contains information both about the local DNA curvature and about the global character of chromosomal supercoiling. The helical periodicity together with A/T tracts of periods ~100–200 bp appear to be involved in the overall packing of bacterial nucleoids and eukaryotic chromosomes [5, 20].

In the above examples, periodicity should be understood in a broad statistical sense, since the underlying periodic patterns have been strongly distorted by random point mutations and insertions/deletions during molecular evolution. Using the complete genomic sequence of *E. coli* and the set of promoter sequences from RegulonDB, we performed a detailed Fourier analysis of periodicities in promoter regions and compared the most characteristic periodicities with those in the whole genome and in random sequences. Particular attention was paid to the helical A- and B-like periodicities with the lengths about B-helix pitch ~10.2–10.5 bp and A-helix pitch ~10.8–11.1 bp as well as to the alternating PuPy periodicities potentially responsible for B→Z transition. We revealed that A- and B-like periodicities coexist in genomic DNA sequences apparently independently and play presumably different regulatory roles. Comparison with experimental data [21–24] indicates that the promoters with the most pronounced periodicities belong to the supercoiling-sensitive genes (SSG). The expression of SSG is regulated by supercoiling as a form of feedback control. In accordance with topological predictions, we found that the relaxation of negative supercoiling downregulates the genes related to the promoters with A- and Z-like periodicities and upregulates the genes related to the promoters with B-like periodicity.

A- and B-like periodicities play probably a multifunctional role. As is known, unlike the B-helical structure of genomic dsDNA, DNA–RNA hybrids form an A-helical structure during transcription [25–27]; also, interaction with regulatory proteins or RNA polymerase may induce a B→A transition in genomic DNA [28, 29]. The corresponding periods of the A-helix, ~10.8–11.1 bp, may be reflected in the sequence periodicities of genomic B-form DNA. The concordance or discordance between A- and B-periodicities may facilitate or hamper the formation of DNA–RNA hybrids and affect the transcription. Thus, the study of underlying periodicities may help elucidate the intricate relationships between structure and function in genomic DNA sequences.

**Materials and methods**

The genomic sequence of *E. coli* K-12 MG1655 and the other genomes used in our study were retrieved from the GenBank (ftp://ftp.ncbi.nih.gov/genbank/genomes). The coding and intergenic regions on the forward and reverse DNA chains were determined according to the GenBank annotations. *E. coli* K-12 transcription start sites and promoter sequences were retrieved from the RegulonDB database release 6.8 (http://regulondb.ccg.unam.mx). For our bioinformatic analysis we developed a set of programs, including two optimized Fourier transform programs that are based upon OpenMP and OpenCL technologies. The visualization of the genomic periodicities was performed with DNAPlotter [30].

**Theory/calculation**

*Fourier transform and statistical criteria*

The latent periodic patterns in DNA sequences can be efficiently displayed by Fourier transform [5, 6, 31–35]. The brief summary below follows earlier publications [31, 35]. Fourier harmonics corresponding to nucleotides of type α (where α is A, C, G, or T) in a sequence of length $L$ are calculated as

$$\rho_\alpha(q_n) = L^{-1/2} \sum_{m=1}^{M} \rho_{m,\alpha} e^{-iq_n m}, \quad q_n = 2\pi n/L, \quad n = 0,1,\ldots,L-1 \qquad (1)$$

Here $\rho_{m,\alpha}$ indicates the position occupied by the nucleotide of type $\alpha$; $\rho_{m,\alpha} = 1$ if the nucleotide of type $\alpha$ occupies the *m*-th site and 0 otherwise. The amplitudes of Fourier harmonics (or structure factors) are expressed as

$$F_{\alpha\alpha}(q_n) = \rho_\alpha(q_n)\rho_\alpha^*(q_n) \qquad (2)$$

where asterisk denotes the complex conjugation. The zeroth harmonics, depending only on the nucleotide composition, do not contain structural information and will be discarded below. The structure factors will always be normalized with respect to the mean spectral values, which are determined by the exact sum rules,

$$f_{\alpha\alpha}(q_n) = F_{\alpha\alpha}(q_n)/\overline{F}_{\alpha\alpha}; \quad \overline{F}_{\alpha\alpha} = N_\alpha(L-N_\alpha)/L(L-1) \qquad (3)$$

where $N_\alpha$ is the total number of nucleotides of type $\alpha$ in a sequence of length *L*. The spectrum of structure factors (2) is symmetrical relative to $q_n = \pi$. Therefore, the spectrum can be restricted to the left half, $q_n \leq \pi$ or $1 \leq n \leq L/2$. The characteristic period and harmonic number are related as $p = L/n$, though generally the periodicities should be identified through sets of equidistant peaks [31, 34, 35].

In random sequences of the same nucleotide composition, the structure factors with different $q_n$ may be considered as independent and obeying Rayleigh statistics

$$\Pr(f > f') = e^{-f'} \qquad (4)$$

universal in spectral analysis. Averaging over *P* random spectra yields in the limit of large *P* Gaussian distribution with the mean and standard deviation given by $<f> = 1, \sigma(f) = 1/\sqrt{P}$. Throughout our paper we use a 5% threshold of statistical significance. The particular criteria were taken from [31, 35]. The pronounced peaks in the genomic spectra should be compared with the singular outbursts in the random spectra by extreme value statistics rather than (4). Following previous studies [17, 18], we searched for significant helical periodicities in the complete genomes within the 9.5–11.3 bp range. We used the 5% criterion that a maximum from harmonics within the corresponding range in a random spectrum may exceed the threshold defined by

$$\Pr = 0.05 = 1 - (1 - e^{-f})^N \tag{5}$$

or $f_{\text{thr}} \approx \ln N + 2.99\ldots$ Here $N$ is the total number of harmonics within the range under investigation, $N = M/9.5 - M/11.3$, and $M$ is the length of complete genome. For $M \approx 10^6$ bp the total number of harmonics is about $N \approx 10^4$.

*Supercoiling and periodicities in DNA sequences*

The chromosome of *E. coli* is packed into a nucleoid structure consisting of 40–50 negatively supercoiled domains [13]. The density ~2.5–5 superturns/kb corresponds approximately to one superturn per Kuhn length characterizing dsDNA bending rigidity (note that Kuhn length is twice as large as persistence length [36]). In dsDNA with fixed ends, the supercoiling tension induces the changes in the helix twist and in the conformational writhe of a strand according to

$$Tw + Wr = Lk = M/p_0 + \tau \equiv M/p_0(1+\sigma) \tag{6}$$

where $Lk$ is the linking number, $\tau$ is the number of superturns in a genome of length $M$, and $p_0$ is the helix pitch in the absence of supercoiling [36, 37]. At equilibrium, a superturn is distributed between twisting and writhing in the proportion $(Tw - M/p_0) : Wr \approx 1 : 2.7$ [36]. For $p_0 \approx 10.2 - 10.5$ and at a density $\pm(2.5-5)$ superturns/kb the shifted pitch would be confined within $\mp 0.1-0.2$ around $p_0$[1], less than the observed shifts in the sequence periodicities [17–19]. The actual redistribution of the superturns *in vivo* depends strongly on the regulatory proteins, gyrase and topoisomerase I.

**Results**

*Averaged promoter spectra*

---

[1] The reversal of signs reflects that positive supercoiling tightens dsDNA and diminishes helix pitch, whereas negative supercoiling unwinds dsDNA and increases helix pitch [17–19].

The complete set comprised 1648 *E. coli* promoters. Some of the retrieved sequences overlapped, so for control we also used subsets of 683 non-overlapping promoters on the forward chain and 699 non-overlapping promoters on the reverse chain. The results proved to be robust with respect to the set. The window length for Fourier analysis was chosen to be 101 bp for the following reasons. (i) This is perhaps the shortest window resolving generic periodicities and ensuring the locality of analysis. (ii) The window is a bit longer than the RNA polymerase contact region. (iii) The window is slightly shorter than the persistence length of dsDNA (~150 bp). (iv) 101 bp is about the mean length of the non-coding regions on dsDNA determined from the coincident non-coding regions on the forward and reverse chains. Similarly to PromEC database (http://margalit.huji.ac.il/promec/), the promoters will be associated with the region (–75, +25) from the transcription start (TS). The ±300 vicinity around TS was also studied with step = 1.

Fourier half-spectra averaged over the promoter sets are shown in Fig. 1. For A and T spectra, the high harmonics with the numbers $n = 1$–$2$ (or periods $p = 101 - 50.5$ nucleotides (nt)) are associated with the variations in A/T composition on the window scale, typical of A/T-rich promoters. The peaks at $n = 5$–$7$ ($p = 20.2$–$14.4$) indicate the spacer-period patterns (determined by the distance between the canonical –35 and –10 elements of the promoter). The harmonics at $n = 9$ ($p = 11.2$) and $n = 10$ ($p = 10.1$) will be associated with A-like and B-like periodicities, respectively. The high peak in the G spectrum at $n = 34$ ($p \approx 3.0$) reveals 3-nt periodicity nearly universal in the protein-coding regions [5, 31, 33, 35, 38]. The spectra in Fig. 1 prove the importance of performing Fourier analysis separately for the nucleotides of each type, because the nucleotides of different types may refer to different structural and functional features in genomic DNA (cf. the particular examples in Introduction).

The averaged A and T spectra are strongly correlated (Pearson correlation coefficient is 0.9). We used the sum $f_{AA}(q_n) + f_{TT}(q_n)$ of structure factors at $n = 9$ or $n = 10$ as an indicator of A- or B-like periodicity. This sum is invariant with respect to complementary inversion of a sequence [34] and is more convenient for comparing the helical periodicities in the complete genome and in the promoter sequences (in the latter case, the promoters on two chains were always compiled as 5´–3´ sequences). The mean and standard deviation for the random sequences are $\langle f_{AA}(q_n) + f_{TT}(q_n)\rangle = 2$ and $\sigma = \sqrt{2}$. Averaging over *P* patterns retains the mean but reduces the standard deviation to $\sigma = \sqrt{2/P}$ (≈ 0.03 for the complete promoter set). The averaged spectra $f_{AA}(q_n) + f_{TT}(q_n)$ in the vicinity of TS are shown in Fig. 2. The largest values for the helical periodicities, 2.6–2.7, exceed the mean by ~20$\sigma$ and are highly statistically significant. Except for the coding periodicity $p \approx 3.0$ ($n = 34$), growing stronger in the region downstream of TS, all other pronounced periodicities peak upstream of TS approximately at the position −125 (measured by the distance between TS and the 5´-end of the 101-nt sliding window). For the helical periodicity, such behavior has been reported previously [9] and implies a higher curvature and a strong impact of this region on the transcription regulation. The ranking of periodicity intensities at −125 derived from Fig. 2 is $p$ = 101 > 11.2 > 50.5 > 10.1 > 33.7 > 14.4 > 25.3 > 16.8 > 20.2.

The promoters and regions upstream of TS are enriched in palindromic repeats typical of enteric bacteria [39, 40] (see Fig. 3 (a)). The palindromic repeats were searched by the following algorithm. For all different words of length $l$ within a given window the non-overlapping complementary counterparts were searched within the same window with the correspondence exceeding 80%. The correspondence 80% is defined by the relative number of substitutions between the perfect and observed complementary counterparts. The palindromic repeats may also be displayed via the high A–T and G–C correlations of the averaged spectra (Fig. 3(b)). The resemblance between corresponding averaged spectra for the promoter set is clearly seen in Fig. 1. The underlying palindromic repeats in the region upstream of TS are A/T-rich and differ from G/C-rich repeats reported previously [39, 40].

We studied also the alternating PuPy periodicities in promoters potentially responsible for B→Z transition. In this case, four-letter DNA sequences were converted into two-letter purine R = (A, G) and pyrimidine Y = (C, T) sequences. Z-like periodicity was searched for via the maximum structure factor $f_{RR}(q_n) = f_{YY}(q_n)$ from the harmonics with $n$ = 48, 49, and 50 ($p \approx 2.0$). The corresponding random mean and standard deviation values are $<\max\{f_j\}_{j=1}^{3}> = 11/6$ and $\sigma = 7/6$ [35]. The promoters with the strongest A-, B-, and Z-like periodicities are presented in Table 1. They were filtered by the criteria $f_{AA}(q_n) + f_{TT}(q_n) > 2 + 4\sqrt{2} \approx 7.66$ for A- and B-like periodicities and max $f_{RR}(q_n) > 11/6 + 3 \times 7/6 \approx 5.33$ for Z-like periodicity. The promoters with strong A- and B-like periodicities reproduced in part those listed in Table 1 of Ref. [4] and filtered by the predicted curvature of dsDNA.

*Comparison of A-, B-, and Z-like periodicities in promoter sequences and in the whole genome of E. coli*

The distributions of A-, B-, and Z-like periodicities in the promoter set were compared with the analogous distributions in the *E. coli* genome and in random sequences. The complete genomic sequence was covered with non-overlapping concatenated windows of length 101 nt and Fourier analysis was performed for each window separately. A-like periodicity was defined by the sum $f_{AA}(q_n) + f_{TT}(q_n)$ with $n$ = 9 corresponding to the period $p$

= 11.2, B-like periodicity was defined by the sum $f_{AA}(q_n) + f_{TT}(q_n)$ with $n = 10$ corresponding to the period $p = 10.1$, whereas Z-like periodicity was defined by the maximum structure factor $f_{RR}(q_n) = f_{YY}(q_n)$ from the harmonics with $n = 48$, 49, and 50 ($p \approx 2.0$). Then, the fractions of windows with the structure factors exceeding given threshold (the relevant current thresholds are shown on X-axes in Fig. 4) were determined. The divergence between such empirical distributions may be conveniently assessed with Kolmogorov-Smirnov criterion. The reference random sequences were obtained by stochastic reshuffling of nucleotides within each window. The results of simulation for the random sequences proved to be in good agreement with the theoretical predictions [35].

Fig. 4 shows the abundance of A- and B-like periodicities in the promoter region over the whole genome: both periodicities are more abundant than in the random sequences. The abundance is the highest at the position –125 from TS. On the contrary, Z-like periodicity is underrepresented in the promoter and genomic sequences as compared with the random ones, but it is more abundant in the promoter region than in the whole genome. The distribution for A-like periodicity goes slightly over the corresponding distribution for B-like periodicity. The correlations between A-, B-, and Z-like periodicities in both the promoter and the genomic sets do not statistically differ from those in the random sequences. The fraction of sequences in which A- and B-like periodicities simultaneously exceed a given threshold approximately coincides with the product of the respective fractions for each periodicity, supporting thus the suggestion about their independence.

The distributions of A-, B-, and Z-like periodicities along the *E. coli* genome are shown in Fig. 5 together with the coding and intergenic regions on the forward and reverse chains. All values were divided by the respective mean plus two standard deviations for the random sequences. Such normalization corresponds to 5% significance level (i.e., among random sequences only 5% would have the periodicity intensities exceeding the normalization value). The observed fractions of the genomic sequences exceeding this threshold were 9.0 and 7.7% for A- and B-like periodicities, respectively. Taking into account the large number of sequences in the genomic set ($P = 45,937$), the corresponding deviations are highly statistically significant.

The more intense are the periodicities, the more vividly they tend to localize in the non-coding regions (see Fig. 6). Previously, the similar but much steeper dependence was reported for the helical periodicity of 11.2 nt in the *E. coli* genome [10]. For A-like periodicity, we observe this effect only at the highest intensities, whereas at lower intensities it appears to be more abundant in the coding rather than non-coding regions.

*Helical periodicities in bacterial and archaeal genomes*

Fourier analysis for nucleotides of different types was applied to the complete genomes: 4 bacterial, 4 archaeal, and 1 eukaryotic. The typical length of genomes $M$ exceeded several megabases. The significant structure factors $f_{\alpha\alpha}(q_n)$ were searched for in the range $n \in (M/11.3, M/9.5)$ by the criterion (5) and collected in Table 2. As the resolution between periods is improved with increasing $M$, the different helical periods in the complete genomes were resolved with high precision. From Table 2 it is evident that pronounced periodicities identified by singular peaks are rather rare events. For the three strongest periodicities in the *Pyrococcus abyssi* genome, the structure factors are somewhat lower than the given threshold. If all pronounced local periodicities shown in Fig. 5 were phased with each other, the heights of the structure factors for the whole genomes would be many orders of magnitude higher than those in Table 2. The net phasing remains still significant. We have not found any significant periodicities near $p = 2.0$ for R–Y spectra in the complete genomic sequences. This means the absence of significant phasing for Z-like periodicities in the whole genomes.

In Refs. [17–19] the most intense periodicities in various genomes were determined by using different methods: mutual information, correlation functions, and Fourier transform. The direct comparison between data in Table 2 and those in Refs. [17–19] needs additional data processing. The key items are related to the proper normalization and statistical criteria. Additionally, the authors used the smoothing of correlation functions [18] or the similar procedure of fitting [17, 19]. Such a smoothing leads to a coarse-graining of initial Fourier spectrum and to some ambiguity in the definition of the most intense periodicities. In our paper we were primarily interested in the polymorphism of helical periodicities and their genetic meaning. All periodicities reported previously as the most intense in Refs. [17–19] are contained among the significant periodicities in our Table 2 up to the coarse-graining uncertainty. The choice of "the most intense" among the significant periodicities needs the proper definition and statistical re-evaluation, if the coarse-grained maximums are substituted instead of singular spectral peaks. Following [17, 18], we assessed the most intense periodicities via the maximum of the sum $f_{AA}(q_n) + f_{TT}(q_n) + f_{CC}(q_n) + f_{GG}(q_n)$ and found the periods: 11.05; 10.91; 11.00; 11.19; 10.85; 9.84; 9.94; 10.62 (the sequence of periods corresponds to the list of bacterial and archaeal genomes in Table 2). The period for *S. cerevisiae* is not included because the maximum in this case is not pronounced. The agreement with previous results is clear (up to the uncertainties related to the normalization and coarse-graining).

**Discussion**

The underlying periodic patterns may play multiple roles in genomic DNA, being specifically recognized and cooperatively bound by regulatory proteins and being involved in the main biological processes such as replication, recombination, and chromatin packing. The stretches with A-, B-, and Z-like periodicities provide the convenient triggers which may be recognized by regulatory proteins and simultaneously be governed by the level and sign of supercoiling. The topological consideration implies that the genes with pronounced A- and Z-like periodicities should be switched on under the enhancement of negative supercoiling, whereas the genes with B-like periodicities should be switched off under the same conditions, and vice versa if the supercoiling is relaxed. In earlier works [21–23], relaxation of the *E. coli* chromosome was shown to upregulate *gyrA* and *gyrB* and downregulate *topA*. As can be seen in Table 1, the promoters related to the genes *gyrB* and *topA* are among the stretches with the strongest B- and A-like periodicities, respectively, while the corresponding genes express in accordance with topological predictions. Later, Peter et al. [24] performed complete screening of supercoiling-sensitive genes (SSG) in *E. coli* with microarray technique. Their screening was based on the significance criteria filtering out the genes with relatively strong irregular variations in expression levels caused by low expression or experimental faults; this procedure underestimates the true number of SSG. Peter et al. [24] identified 106 genes with expression increased under relaxation of negative supercoiling and 200 genes with decreased expression. We found the following promoters in Table 1 corresponding to their significant SSG: *topA* and *agaR* (pronounced A- and Z-like periodicities, respectively; the corresponding genes downregulated under relaxation of supercoiling), *gyrB* and *dnaA* (pronounced B-like periodicities; upregulated under relaxation of supercoiling). The *gloA* promoter turns out to be the only exception (B-like periodicity; downregulated instead of upregulated under relaxation of supercoiling). The comparison with RegulonDB revealed 43 promoters for 26 upregulated SSG and 107 promoters for 86 downregulated SSG. The spectra $f_{AA}(q_n) + f_{TT}(q_n)$ averaged over the first set of promoters show the highest helical periodicity at $n = 10$, $p = 10.1$, whereas for the second set the highest helical periodicity appears to be at $n = 9$, $p = 11.2$. The complete analysis of the results by Peter et al. [24] will be presented separately. The greater total number of downregulated genes agrees also with the higher percentage of stretches with the pronounced A-like periodicities in the *E. coli* genome. Except for the highest intensities, the A-like and the B-like periodicities are quite differently distributed between the coding and the non-coding regions (Fig. 6). This also indicates their different regulatory functions.

The data in Table 2 reveal the clusters of periodicities of B-like type $p \approx 10.2–10.4$, A-like $p \approx 10.8–11.1$, and the third cluster with $p \approx 9.6–9.9$. Such grouping agrees with the "twin-supercoiled-domain" model of transcription [41–44]. The processive motion of RNA polymerase induces the positive supercoiling downstream and negative supercoiling upstream of the transcribed DNA stretch. The corresponding positive and negative supercoiling is relaxed by gyrase and topo I, respectively, yet a lag should be expected in this relaxation process. If the positive and negative superturns induced by the processive motion of RNA polymerase will, first, be redistributed by the elastic tension over a distance about persistence length, the shifts in the helix pitch would be about the observed periodicities, since the local density of superturns during transcription is higher than that of discussed in Theory/calculation. The universality of this mechanism implies the coexistence of helical periodicities ~10, ~10.5, and ~11 bp in genomic sequences. The global chromosomal supercoiling may shift their relative strength toward ~10 or ~11 bp depending on the sign of supercoiling, but should generally retain the significance of all three periods. These conclusions are in line with Table 2.

The packing of dsDNA into the nucleosomal complex in eukaryotes needs about 1.5 negative superturns per nucleosome. The elastic tension during winding is compensated by the interaction with histone octamer, and dsDNA in a nucleosome is in a mechanically relaxed state. In accordance with this conclusion, the underlying periodicity in *Saccharomyces cerevisiae* chromosome IV exhibits a unique singular peak $p = 10.39$, $f_{AA} = 13.38$ in the adenine spectrum, coinciding with the B-helix pitch. However, even in this case the two next ranked periodicities are $p = 9.57$, $f_{AA} = 12.01$ and $p = 11.30$, $f_{AA} = 11.90$.

The sites with the most pronounced periodic patterns appear often to be the strongest regulators [1, 2, 16, 45, 46] and are worth being investigated experimentally. Promoter-like elements are quite amply scattered throughout *E. coli* genome [47], but their ability to initiate transcription depends strongly on the structural surroundings. The prevalence of the main generic periodicities in the promoter set over the corresponding genomic periodicities (cf. Fig. 4) suggests their signal significance. Our study reveals that region upstream of TS contains both the pronounced helical periodicity [9] and actually all other longer periodicities up to the largest tested scale of 101 bp (Fig. 2). Besides the principal interest, the study of underlying periodic patterns may be integrated in the general program for *in silico* search and *in vivo* assessment of the putative regulatory sites.

**Acknowledgements**

The authors are grateful to A.V. Galkin for stimulating discussions and for editing the text. This work was supported by the Molecular and Cellular Biology Program of the Presidium of the Russian Academy of Sciences.

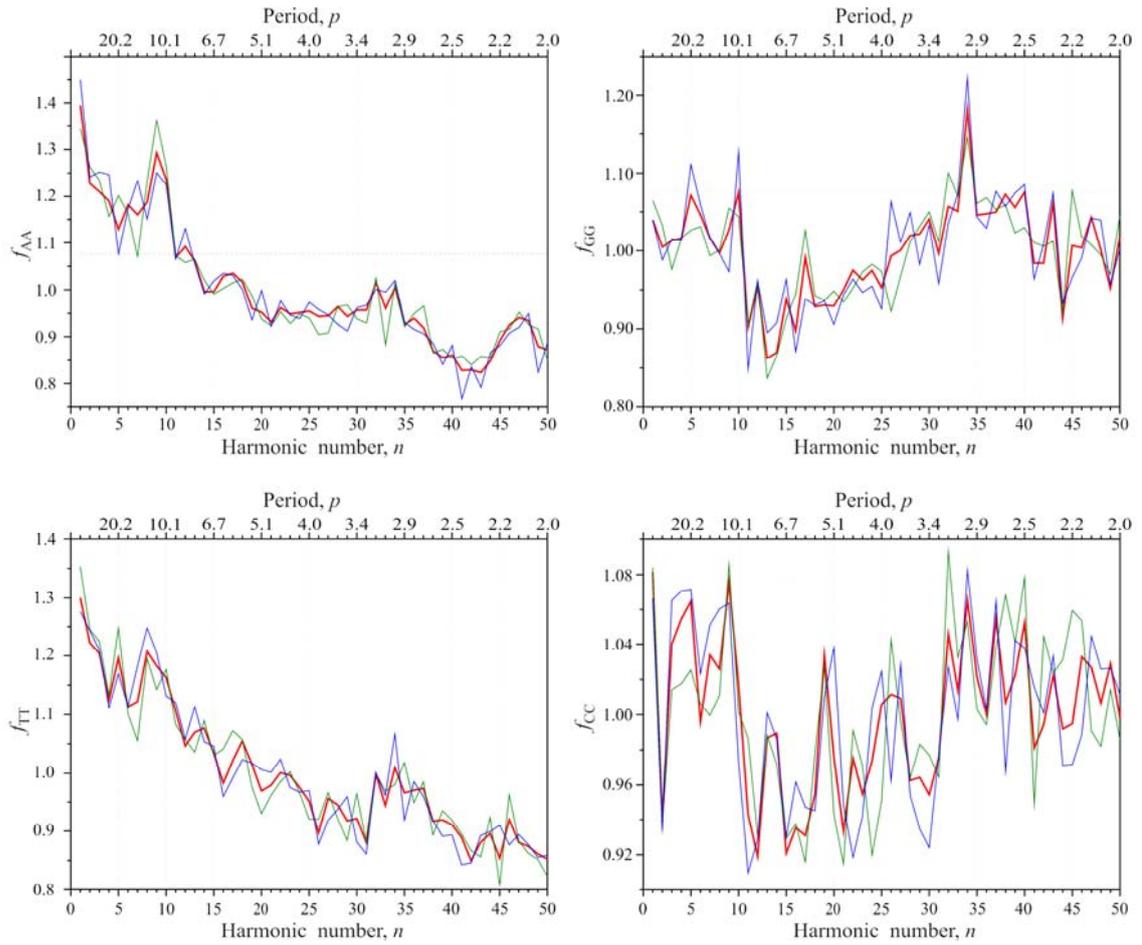

**Fig. 1.** Fourier half-spectra for the structure factors averaged over the sets of *E. coli* promoters. Red line, all promoters; green line, non-overlapping promoters on the forward chain; blue line, non-overlapping promoters on the reverse chain. The horizontal line corresponds to 5% probability that a maximum from harmonics in an averaged half-spectrum for the random sequences of the same nucleotide composition as the counterpart promoters exceeds this level. The significance level corresponds to the complete set of promoters; the variations of this level for two other sets are minor.

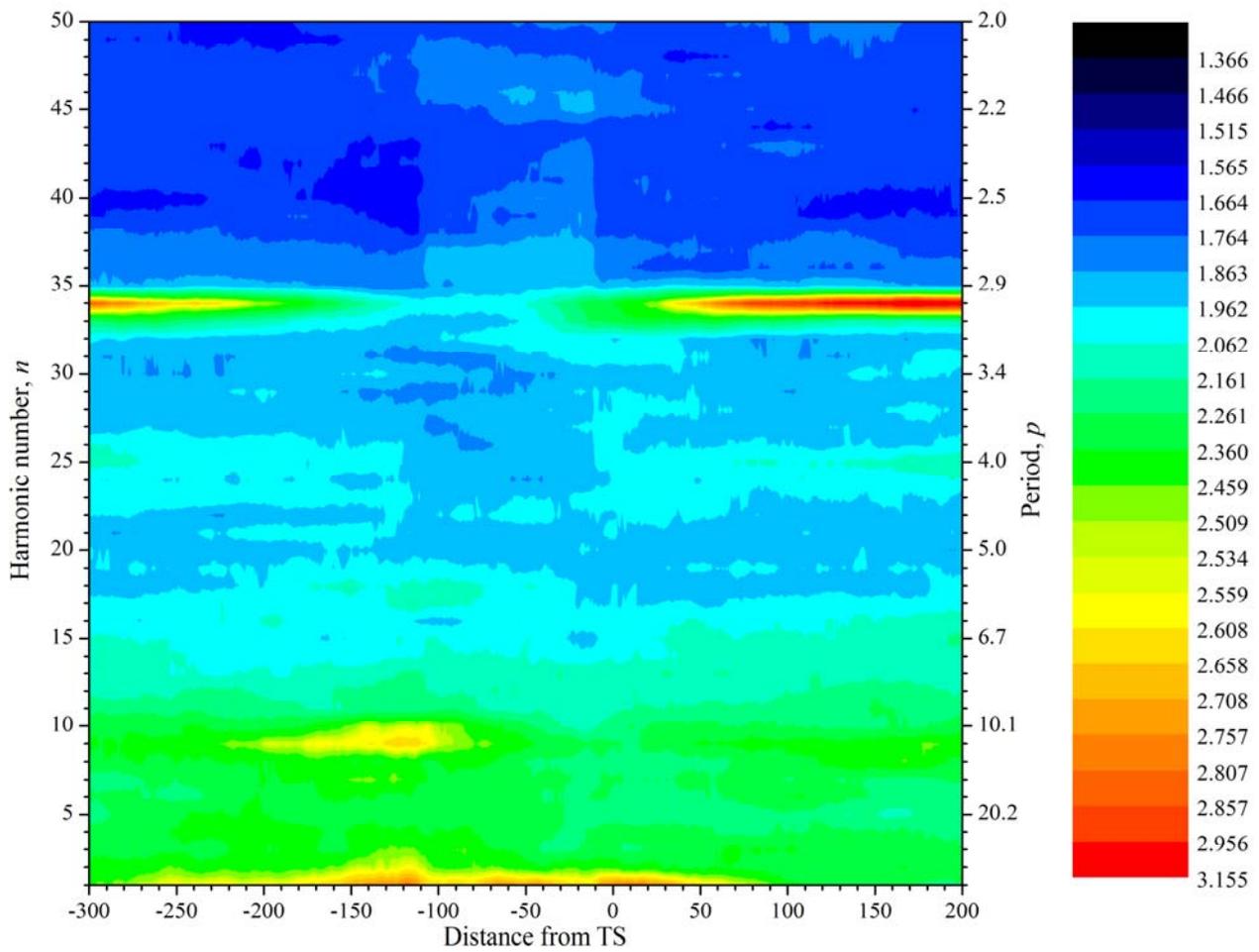

**Fig. 2.** The averaged spectra for the sum $f_{AA}(q_n) + f_{TT}(q_n)$ of structure factors in the vicinity of promoter region. The position from the transcription start (TS) is measured by the distance from 5´-end of a 101-nt sliding window. The averaging is performed over all sliding windows at given position from TS. Scale on the right corresponds to the sum $f_{AA}(q_n) + f_{TT}(q_n)$.

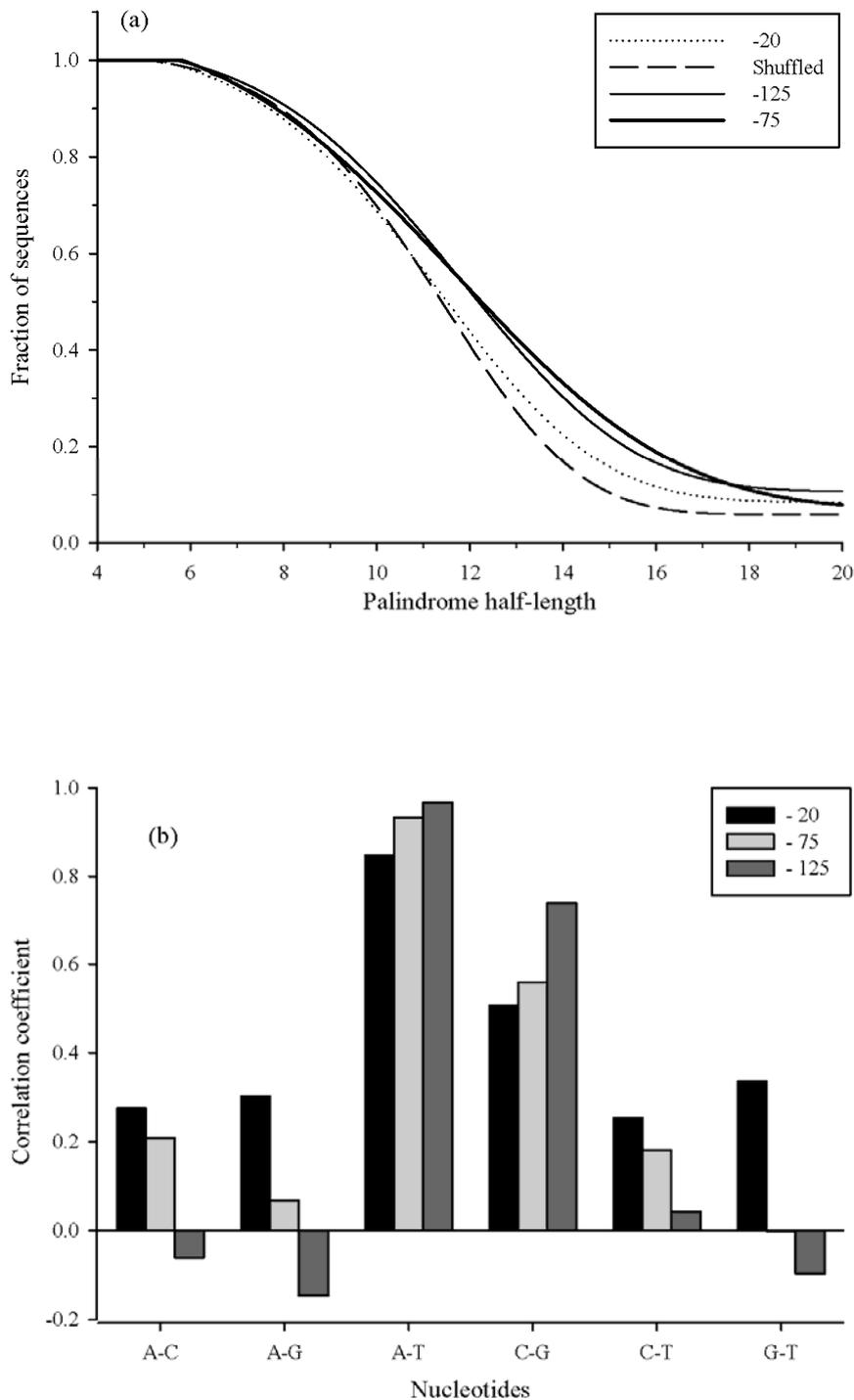

**Fig. 3.** (a) The distribution of palindromic repeats in the sets of 101-nt windows at the different positions from TS. The location is defined by distance between 5′-end of a window and TS. The promoter set is located at −75. The palindromic repeats were searched as follows. For all different words of length $l$ within a given window the non-overlapping complementary counterparts were searched within the same window with the correspondence exceeding 80%. The reference distribution for the random sequences was obtained by stochastic reshuffling of nucleotides in the promoter set. (b) Pearson correlation coefficients between the averaged structure factor spectra at different positions from TS. Only A–T and G–C correlations are statistically significant. The statistical significance of correlation coefficients may be assessed either by simulations or by the criteria presented in Ref. [35].

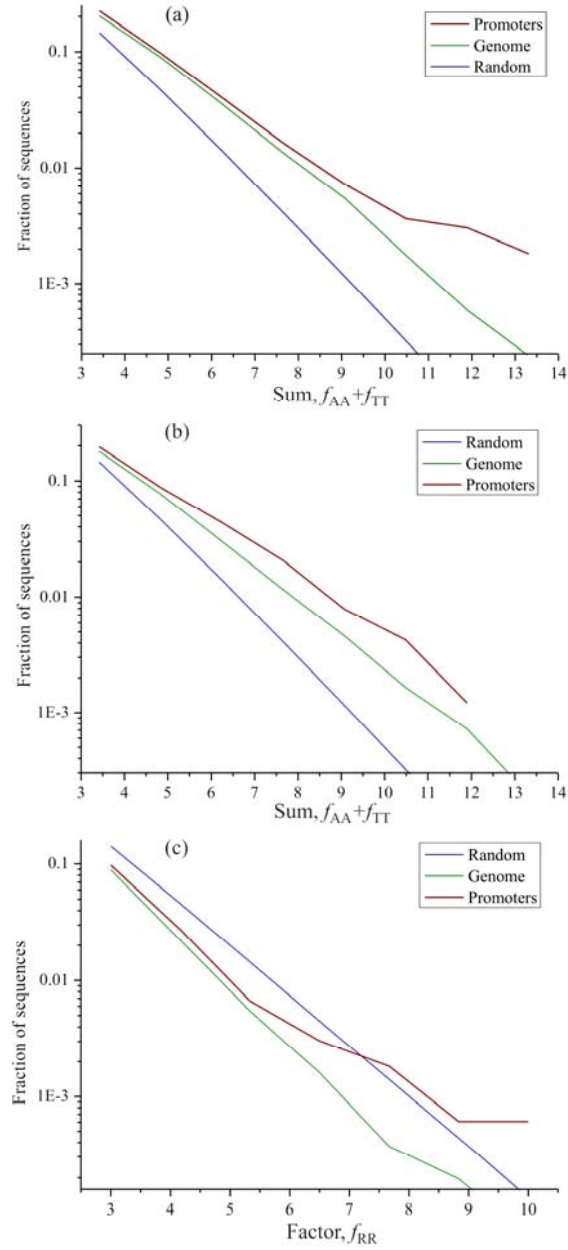

**Fig. 4.** The distributions of the structure factor heights for (a) A-, (b) B-, and (c) Z-like periodicities in the promoter sequences and in the whole genome of *E. coli*. A-like periodicity was defined by the sum $f_{AA}(q_n) + f_{TT}(q_n)$ with $n = 9$ corresponding to the period $p = 11.2$, B-like periodicity was defined by the sum $f_{AA}(q_n) + f_{TT}(q_n)$ with $n = 10$ corresponding to the period $p = 10.1$, whereas Z-like periodicity was defined by the maximum structure factor $f_{RR}(q_n) = f_{YY}(q_n)$ from the harmonics with $n = 48, 49$, and $50$ ($p \approx 2.0$). The reference distributions for the random sequences were obtained by reshuffling nucleotides in the genomic windows and calculating corresponding Fourier spectra after reshuffling. Y-axis: the fraction of windows in which the height of structure factors exceeds given threshold; X-axis: current threshold for the height of structure factors.

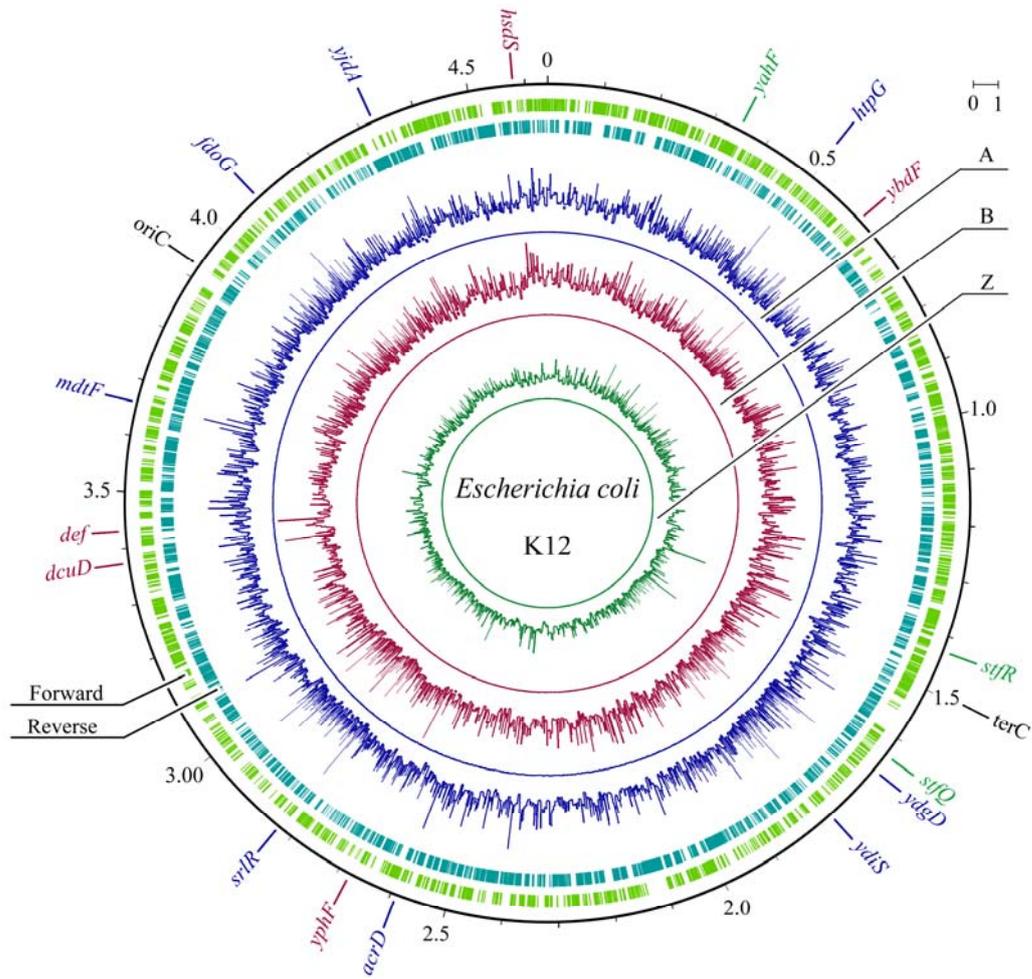

**Fig. 5.** From outside to inside: The distribution of the coding and intergenic regions on the forward and reverse chains in the *E. coli* genome; the distribution of $f_{AA}(q_n) + f_{TT}(q_n)$ structure factors for A-like and B-like periodicities in the genomic nucleotide sequence; and the distribution of $f_{RR}(q_n) = f_{YY}(q_n)$ structure factors for Z-like periodicity. To unify the comparison, all structure factors are normalized to 5% significance level for the random counterparts. The normalized height for the structure factors is shown at the top right corner. The genomic length scale in megabases and the particular genes containing stretches with the most pronounced periodicities are shown in the outer region.

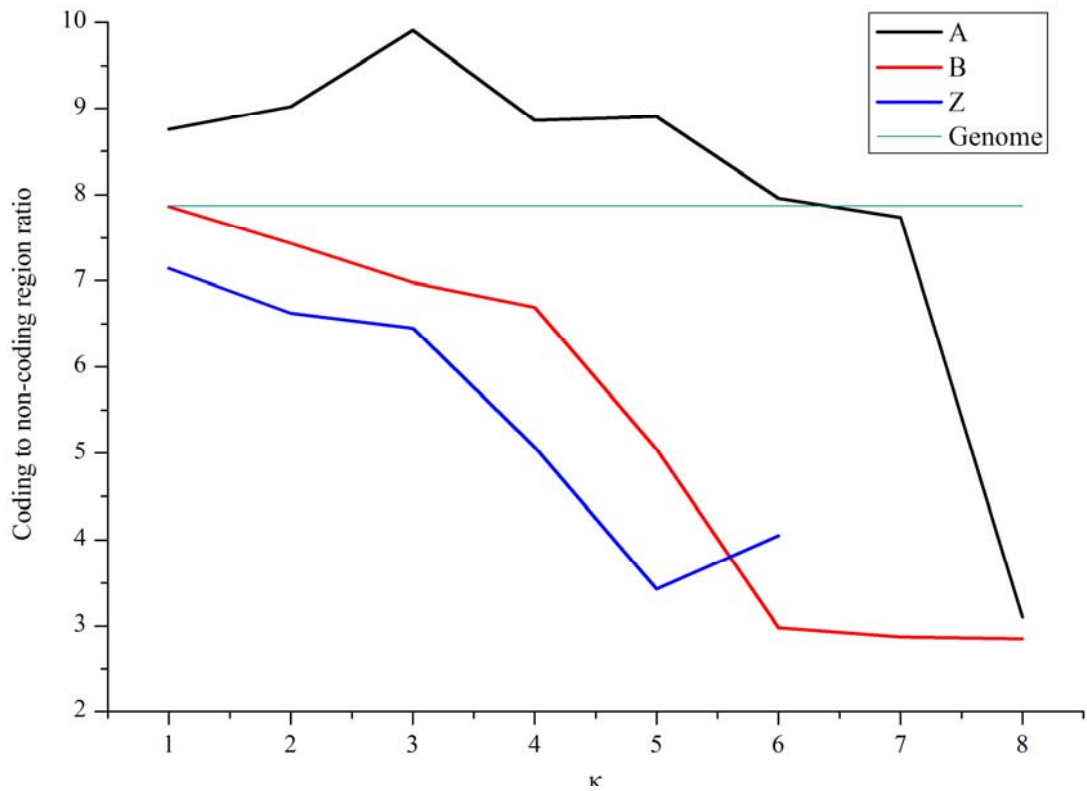

**Fig. 6.** The distributions of stretches with A-, B-, and Z-like periodicities over coding and non-coding regions in the *E. coli* genome. The non-coding regions on dsDNA are determined from the coincident non-coding regions on the forward and reverse chains. The horizontal line corresponds to the ratio of the total length of coding regions to that of non-coding ones in the whole genome. To unify the comparison, the periodicity intensities on X-axis are defined in terms of mean + *k* standard deviations for the random counterparts.

**Table 1**

*Escherichia coli* promoters with the strongest A-, B-, and Z-like periodicities

| Periodicity | Promoter |
|---|---|
| A-like | *yhjH* P, *katE* P, *talA* P1, *cydA* P5, *cytR* P, *fliA* P1, *ilvL* P2, *kdpF* P, *modA* P, *nuoA* P2, *ompX* P1, *purM* P, *pyrD* P, *trpL* P, *udp* P, *ysgA* P, *copA* P, *glnB* P, *kup* P2, *potF* P2, *ravA* P, *sspA* P, *tktA* P1, *topA* P3, *tyrS* P2, *xylR* P, *ycgE* P |
| B-like | *ppdA* P, *ldhA* P, *aldB* P, *carA* P1, *cspD* P, *ftsK* P2, *gloA* P, *gyrB* P, *hfq* P2, *lys* P, *molR_1* P, *mraZ* P, *pck* P, *rpsU* P2, *rpsU* P3, *ruvC* P, *symE* P, *valS* P1, *valS* P2, *ychF* P, *yjjQ* P, *yrbG* P, *acp* P, *cyaA* P1, *dnaA* P1, *dnaA* P2, *eda* P1, *nlpD* P1, *smf* P, *tolC* P2, *trxB* P, *ydgA* P, *csiE* P, *osmY* P |
| Z-like | *rpmE* P, *lsrA* P, *yhdW* P, *acrA* P, *agaR* P, *cmk* P, *gnd* P, *chpR* P1, *chpR* P2, *cysK* P2, *rplT* P |

**Table 2**

Helical periodicities for nucleotides of different types in complete genomes

| Genome | Period, nt | | | |
|---|---|---|---|---|
| | A | T | C | G |
| *Escherichia coli* | – | 9.89 (14.64) | – | – |
| | | 10.22 (14.50) | | |
| | | 10.99 (14.25) | | |
| *Bacillus subtilis* | 9.88 (14.50) | 10.72 (17.54) | – | 10.91 (14.15) |
| *Mycobacterium tuberculosis* | 11.00 (14.52) | – | – | – |
| | 10.57 (14.27) | | | |
| *Helicobacter pylori* | 11.19 (16.19) | 11.00 (17.84) | – | 10.66 (13.28) |
| | 11.16 (15.30) | 10.93 (17.41) | | |
| | 10.92 (14.37) | 11.19 (15.16) | | |
| | 10.63 (14.21) | 11.24 (13.94) | | |
| | 11.19 (14.16) | 11.00 (13.76) | | |
| | 10.85 (14.04) | 10.53 (13.74) | | |
| | 10.37 (13.96) | | | |
| | 11.06 (13.25) | | | |
| *Pyrococcus horikoshii* | – | 11.08 (15.11) | 10.76 (13.39) | – |
| | | 9.54 (13.27) | | |
| *Methanocaldococcus jannaschii* | 10.60 (14.89) | 10.19 (12.96) | – | – |
| *Archaeoglobus fulgidus* | 10.35 (15.42) | 9.69 (18.54) | – | – |
| | 9.54 (14.93) | 10.67 (15.85) | | |
| | 10.04 (13.73) | 11.25 (15.62) | | |
| | 10.47 (13.58) | 9.94 (15.06) | | |
| | 9.94 (13.41) | | | |
| *Pyrococcus abyssi* | 10.62 (12.80) | – | – | 10.85 (12.91) |
| | | | | 10.74 (12.83) |
| *Saccharomyces cerevisiae* chr. IV | 10.39 (13.38) | – | – | – |

The first number in the columns denotes period; the number in the parentheses denotes the structure factor height. The observed periodicities are ordered by the structure factor height. The first four genomes belong to Bacteria, the next four to Archaea, and the last one to Eukarya.